\documentclass[useAMS,usenatbib]{mn2e}

\usepackage[psamsfonts]{amssymb}
\usepackage[dvips]{graphicx}
\usepackage{amsmath,alltt}                                                                                                          
\usepackage{multirow}
\usepackage{rotating}
\usepackage{lscape}
\title[The Stellar Mass Functions of Globular Clusters]{Modelling the Observed Stellar Mass Function and its Radial Variation in Galactic Globular Clusters}
\author[Webb,J.J. et al.]{Jeremy J. Webb$^{1}$, Enrico Vesperini$^{1}$, Emanuele Dalessandro$^{2,3}$, Giacomo Beccari$^{4}$, \and Francesco R. Ferraro$^3$, Barbara Lanzoni$^3$ \\  
\thanks{E-mail: jerjwebb@iu.edu (JW), evesperi@indiana.edu (EV)} \\
$^1$ Department of Astronomy, Indiana University, Swain West, 727 E. 3rd Street, IN 47405 Bloomington, USA \\
$^2$ INAF Osservatorio Astronomico di Bologna, Via Ranzani 1, Bologna, Italy \\
$^3$ Dept. of Physics and Astronomy, University of Bologna, Viale Berti Pichat, 6/2,
Bologna, Italy \\
$^4$ European Southern Observatory, Karl-Schwarzschild-Strasse 2, D-85748 Garching bei Munchen, Germany}

\begin{document}

\pagerange{\pageref{firstpage}--\pageref{lastpage}} \pubyear{2017}

\maketitle

\label{firstpage}

\begin{abstract}

We measure how the slope $\alpha$ of the stellar mass function (MF) changes as a function of clustercentric distance $r$ in five Galactic globular clusters and compare $\alpha(r)$ to predictions from direct $N$-body star cluster simulations. Theoretical studies predict that $\alpha(r)$ (which traces the degree of mass segregation in a cluster) should steepen with time as a cluster undergoes two-body relaxation and that the amount by which the global MF can evolve from its initial state due to stellar escape is directly linked to $\alpha(r)$. We find that the amount of mass segregation in M10, NGC 6218, and NGC 6981 is consistent with their dynamical ages, but only the global MF of M10 is consistent with its degree of mass segregation as well. NGC 5466 and NGC 6101 on the other hand appear to be less segregated than their dynamical ages would indicate. Furthermore, despite the fact that the escape rate of stars in non-segregated clusters is independent of stellar mass, both NGC 5466 and NGC 6101 have near-flat MFs. We discuss various mechanisms which could produce non-segregated clusters with near-flat MFs, including higher mass-loss rates and black hole retention, but argue that for some clusters (NGC 5466 and NGC 6101) explaining the present-day properties might require either a non-universal IMF or a much more complex dynamical history.

\end{abstract}


\section{Introduction} \label{intro}

For the majority of a globular cluster's lifetime, two-body relaxation is the dominant mechanism that drives its evolution. As stars in a cluster undergo repeated two-body interactions, high-mass stars transfer kinetic energy to low-mass stars and fall inwards while energized low-mass stars migrate outwards \citep[e.g.][]{heggie03}. The rate at which a given cluster undergoes mass segregation depends on its mass and size, as compact lower-mass clusters have shorter relaxation times than extended high-mass clusters.

Assuming a cluster forms without any primordial mass segregation, the stellar mass function (MF) at different clustercentric radii $r$ will initially be the same as the global MF. Stellar evolution will quickly alter the high-mass end of the global MF, but, assuming high-mass stars share the same initial spatial distribution as low-mass stars the MF will continue to be independent of distance from the cluster centre. Therefore, star loss occurring during the cluster's early evolution will not alter the global MF (Webb \& Vesperini 2016 (hereafter WV16), Balbinot \& Gieles 2017). Only later, after the cluster's evolution has been significantly affected by two-body relaxation, can the low-mass ($m < 0.8 M_{\odot}$) end of the MF develop a radial dependence. As higher-mass stars fall inwards and lower-mass stars migrate outwards, the inner MF will become top heavy and its slope $\alpha$ (hereafter we assume a power-law function for the stellar mass function, $dN/dm \propto m^{-\alpha}$, and we refer to $\alpha$ as the slope of the mass function) will start increasing (become less negative) with time while $\alpha$ in the outer regions of the cluster will decrease (become more negative). Hence over time, a radial gradient $\alpha(r)$ will develop and get steeper as the cluster evolves. Only once a radial gradient in $\alpha(r)$ develops will the slope of the global MF evolve from its initial value as the loss of stars will preferentially affect low-mass stars.

Several wide-field studies of Galactic globular clusters have attempted to measure $\alpha(r)$. These studies are forced to combine multiple fields of view of a given cluster, sometimes with different instruments, in order to measure $\alpha(r)$ over a wide enough range in projected clustercentric distance. One of the first such studies was of M10 \citep{beccari10}, where a clear radial dependence in $\alpha$  consistent with the effects of mass segregation was observed. Since then, wide field studies of Pal 4 \citep{frank12},  Pal 14 \citep{frank14}, NGC 5466 \citep{beccari15}, and 47 Tuc \citep{zhang15} have also found evidence of $\alpha$ depending on clustercentric distance. The degree of radial variation in $\alpha$ differs from cluster to cluster, indicating the rate at which each cluster segregates is different. A wide-field study of NGC 6101 \citep{dalessandro15}, on the other hand, found that $\alpha$ remained almost constant with clustercentric distance, suggesting that the radial distribution of stars in the range of masses the authors explored has not been affected by mass segregation.

In a recent study, \citet{webb16} used $N$-body simulations of star clusters to study how the evolution of $\alpha(r)$ depends on a cluster's initial conditions and the external tidal field it experiences. The authors traced radial variation in the MF with the parameter $\delta_\alpha = \frac{d\alpha(r)}{d(ln\frac{r}{r_m})}$, where $r_m$ is the cluster's half-mass radius, and found that $\delta_\alpha$ expectedly decreases with time (i.e. the gradient becomes stronger) as a cluster relaxes and undergoes mass segregation.  \citet{webb16} also found that the evolution of the slope of the global MF $\alpha_G$ was strongly correlated with that of $\delta_\alpha$, as $\alpha_G$ evolves slowly as a function of mass lost if stars escape the cluster when the cluster is dynamically young ($\delta_\alpha$ is near zero) but evolves more rapidly as the cluster ages and $\delta_\alpha$ decreases; the evolution of mass segregation and the flattening of the global MF are different manifestations of the effects of two-body relaxation and the escape of stars from the cluster. Hence the evolution of $\delta_\alpha$ and $\alpha_G$ are closely linked to each other.

In this study, we directly compare the $N$-body simulations in \citet{webb16} to observations of M10, NGC 5466, NGC 6101, NGC 6218, and NGC 6981 in order to determine if their measured global MF is consistent with their $\delta_\alpha$  and dynamical age. Measurements of $\delta_\alpha$ for all five clusters are done using archive images. For each comparison, only model stars that are within the same field of view and mass range as the observed datasets are considered to remove any dependence that $\delta_\alpha$ or $\alpha_G$ may have on these factors. The five observational datasets used in this study are introduced in Section \ref{s_observations}, while the suite of $N$-body simulations that we compare them to are discussed in Section \ref{s_nbody}. In Section \ref{s_results} we calculate the dynamical age, $\delta_\alpha$ and $\alpha_G$ of each observed cluster and compare them to our simulations. Finally, we discuss and summarize the comparisons in Sections \ref{s_discussion} and \ref{s_conclusion}.

\section{Observations} \label{s_observations}

In the following sub-sections, we discuss the five observational datasets over which we calculate $\delta_\alpha$. Each of the datasets has been corrected for contamination, using stars located beyond the tidal radius of the cluster (when possible) or the Besancon model simulation. The completeness level of stars of a given mass has been estimated as a function of clustercentric distance based on artificial star experiments. For detailed information regarding decontamination, completeness estimates, how stellar candidates were selected and how their magnitudes and masses have been determined, the original publications referenced in each section should be consulted.

\subsection{M10}

Initially studied by \citet{beccari10}, the M10 dataset consists of two separate fields of view that were part of GO-10775 (PI: A. Sarajedini). The inner 120'' of M10 were imaged using the Hubble Space Telescope (HST) Advanced Camera for Surveys (ACS) with F606W and F816W filters. Given that M10 has a $r_m$ of 108.6'' \citep{harris96}, the ACS data allows for $\alpha(r)$ to be measured out to 1.1 $r_m$. The radial region between 145'' and 318'' was partially imaged using HST's Wide Field Planetary Camera 2 (WFPC2) with the F606W and F814W filters under Prop: 6113 (PI: Paresce), extending $\alpha(r)$ out to 2.9 $r_m$. 

To calculate $\alpha(r)$, we split the ACS data into four radial bins each containing the same number of stars and the WFPC2 data into two radial bins each containing the same number of stars. The radius associated with each radial bin is the mean radius of all stars in the bin and each radial bin has been corrected for completeness using the estimates from \citet{beccari10}. We elected to measure $\alpha$ in each radial bin for stars between $0.3 M_{\odot}$ and $0.8 M_{\odot}$ as the completeness over this mass range is greater than $50\%$ in each radial bin. It should be noted that while completeness might be over $50\%$ for a wider mass range in a given radial bin, we are forced to use the mass range for which this is true in all radial bins so $\alpha(r)$ is determined using the same mass range over the entire radial extension of the cluster.

\subsection{NGC 5466}

For NGC 5466, a combination of HST and ground-based images originally presented in \citet{beccari13} were used by \citet{beccari15} to measure $\alpha(r)$. HST ACS images in the F606W and F814W bands cover out to 120'' or 0.56 $r_m$ \citep{miocchi13}. Large Binocular Camera (LBC) images in the B and V of stars between 120'' and the cluster's tidal radius at 1580'' (7.4 $r_m$) allows for $\delta_\alpha$ to be measured over a wide radial range. 

Only two radial bins could be used to measure $\alpha(r)$, with the ACS dataset serving as the inner radial bin and the LBC dataset beyond 400'' serving as the outer radial bin. Breaking up the ACS data into multiple radial bins would result in measuring $\alpha(r)$ across the cluster's core radius. With respect to the LBC data, within 400'' there is significant crowding and a completeness level less than $50\%$ for stars with masses less than $0.55 M_{\odot}$, which would make any measurement of $\delta_\alpha$ unreliable. However for stars beyond 400'' where crowding is less important, experiments by \citet{beccari13} find that completeness levels are above $50\%$ for stars between $0.4 M_{\odot}$ and $0.8 M_{\odot}$. Therefore $\alpha$ was measured in each radial bin for stars with masses between $0.4 M_{\odot}$ and $0.8 M_{\odot}$.

\subsection{NGC 6101}

To study $\alpha(r)$ in NGC 6101, \citet{dalessandro15} used a combination of archive HST images originally published by \citet{sarajedini07} and FORS2 images taken with the Very Large Telescope (Prop ID: 091.D-0562; PI:Dalessandro). The HST data, consisting of ACS/Wide Field Camera (WFC) images in the F606W and F814W bands, covers the innermost 120'' or 0.9 $r_m$ (assuming $r_m =$ 128.2'' \citep{dalessandro15}). The FORS2 data, in both $V_{HIGH}$ and $I_{BESSEL}$, partially cover between 150'' and 890'' (1.2 - 6.9 $r_m$). Each of the two datasets were then split into 3 radial bins containing an equal number of stars in order to measure $\alpha(r)$. Using stars between $0.5 M_{\odot}$ and $0.8 M_{\odot}$ ensured that the completeness in each radial bin was above $50\%$. 

\subsection{NGC 6218}

Our study of NGC 6218 also makes use of ACS and FORS2 data (Prop ID: 093.D-0228, PI: Dalessandro), originally presented in \citet{sollima17}. The ACS field of view encompasses the inner 100'' ($0.94 r_m$) while the FORS2 dataset partially covers stars between 150'' and 1500'' (1.4 - 14.1 $r_m$). Since the FORS2 dataset goes beyond the tidal radius of NGC 6218 (1037''), we only consider stars within the tidal radius to calculate $\alpha(r)$. Similar to NGC 6101, each dataset was split into three radial bins which contained the same number of stars. Due to the limited FORS2 dataset, the mass range over which completeness was over $50\%$ in each radial bin was only $0.55-0.8 M_{\odot}$. 

\subsection{NGC 6981}

Similar to NGC 6218, ACS and FORS2 data (Prop ID: 093.D-0228, PI: Dalessandro) from \citet{sollima17} was used to perform a wide-field study of NGC 6981. However since NGC 6981 is much farther away than NGC 6218, the ACS dataset field of view of $\sim 100''$ corresponds to covering a radial range out to $1.9 r_m$. The FORS2 dataset, which partially covers the radial region between 95'' and 840'', goes well beyond the cluster's tidal radius of 447.6''. Therefore again, similar to NGC 6218, only stars within $r_t$ are used to measure $\alpha(r)$. Just like NGC 6218, the two datasets were split into three radial bins each containing the same number of stars and only stars with masses between $0.55-0.8 M_{\odot}$ yielded completeness levels over $50\%$ in each radial bin. 

\section{N-body models} \label{s_nbody}

In \citet{webb16}, the evolution of $\delta_\alpha$ was studied for a large suite of $N$-body simulations that spanned a wide range of initial conditions. For the purposes of this study, we will mainly focus on two models that have different initial sizes with the understanding that different values of $\delta_\alpha$ can be reached by adjusting the model cluster's initial size, mass, black hole retention fraction, and orbit. 

Model clusters were evolved for 12 Gyr using the direct $N$-body code NBODY6 \citep{aarseth03}. We considered the evolution of models initially containing 100,000 stars with initial half-mass radii $r_{m,i}$ of 1.1 pc and 6 pc. The initial radial profile of both models was set equal to a Plummer density profile \citep{plummer11} out to 10 $r_{m,i}$. Individual stellar masses between 0.1 and 50.0 $M_{\odot}$ were generated using a \citet{kroupa93} initial mass function IMF (\citet{webb16} demonstrated that using a \citet{kroupa01} IMF yields the same evolution in $\delta_\alpha$ for stars with $m<0.8 M_\odot$) and the subsequent stellar evolution of each star follows \citet{hurley00} assuming a metallicity of $Z=0.001$ ([$\frac{Fe}{H}]=-1.3$). All of our simulations began with no primordial binaries. In cases where binary stars form, their evolution follows Hurley et al. (2002); WV16 found that unresolved binaries have a negligible impact on the calculation of $\delta_\alpha$ or $\alpha_G$.

In order to include the effects of an external tidal field, model clusters were placed in a Milky Way-like potential consisting of a point-mass bulge ($1.5 \times 10^{10} M_{\odot}$), a \citet{miyamoto75} disk ($5 \times 10^{10} M_{\odot}$, a=4.5 kpc, b=0.5 kpc), and a logarithmic halo that is scaled in order to force a circular velocity of 220 km/s at a galactocentric distance $R_{gc}$ of 8.5 kpc \citep{xue08}. Both model clusters have a circular orbit at 6 kpc.

\section{Results}\label{s_results}

\subsection{Radial Variations in Observed Stellar Mass Functions}\label{s_obsresults}

Within a given radial bin, each containing the same number of objects, stars were separated into 10 mass bins. Allowing for a variable bin size minimizes any bias associated with radially binning the data \citep{maiz05}. The mass $m$ associated with each bin was set equal to the mean mass of stars in the bin. $\alpha$ was then set equal to the slope of the line of best fit to a plot of $\log$($\frac{dN}{dm}$) versus $\log(m)$ found using linear regression, where dN is the total number of stars in the mass bin and dm is the width of the bin. Hence the MF has the form:

\begin{equation}
\frac{dN}{dm}=m^{-\alpha}
\end{equation}

\begin{figure}
\centering
\includegraphics[width=\columnwidth]{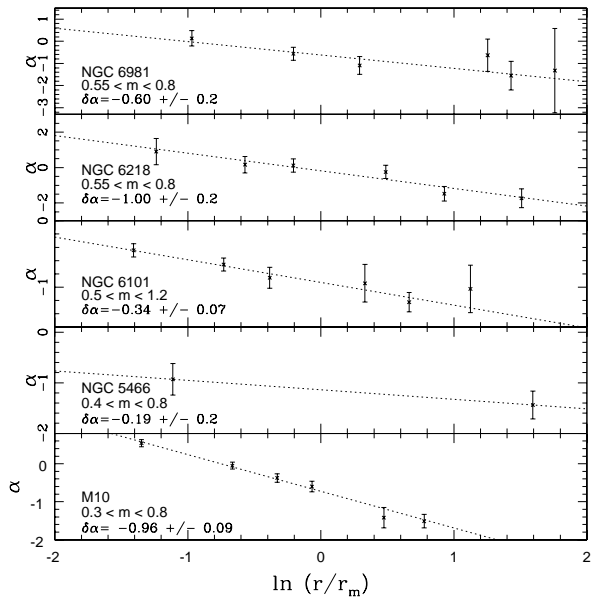}
\caption{Slope of the stellar mass function $\alpha$ versus the natural logarithm of clustercentric radius for globular clusters M10, NGC 5466, NGC 6101, NGC 6218, and NGC 6981. Clustercentric radii have been scaled by each clusters half-mass radius $r_m$. Dotted lines represent the line of best fit to the data, with its slope $\delta_\alpha$ and the mass range over which $\alpha$ was measured noted in each panel. 
}
\label{fig:obs}
\end{figure}

In Figure \ref{fig:obs} we present the radial variation in $\alpha$ as a function of $\ln (\frac{r}{r_m})$ for each observed cluster. The slope of the line of best fit, which quantifies the radial variation of the MF, is referred to as $\delta_\alpha = \frac{d\alpha(r)}{d(ln\frac{r}{r_m})}$. When calculating $\delta_\alpha$, the fit is weighted by the error bars presented in Figure \ref{fig:obs}, which correspond to the linear regression fit to each $\alpha$.

All of the clusters in Figure \ref{fig:obs} clearly show evidence of mass segregation. To compare the degree of mass segregation in each cluster to its dynamical age, we need to know each cluster's age and half mass relaxation time $t_{rh}$. From \citet{spitzer71}, $t_{rh}$ is calculated as:

\begin{equation}
t_{rh}=2.054 \times 10^6 yr \frac{M^{\frac{1}{2}}}{\bar{m}} \frac{r_h^{\frac{3}{2}}}{\ln(0.4 \frac{M}{\bar{m}})}
\end{equation}

where M is the cluster's mass, $r_h$ is the projected half-light radius, and $\bar{m}$ is the mean stellar mass (assumed to be $\frac{1}{3} M_\odot$ \citep{harris96}). Globular cluster ages are taken from \citet{forbes10}, who calculates absolute ages from the relative ages determined by \citet{marinfranch09} and assuming a normalisation of 12.8 Gyr (the mean absolute age of metal-poor globular clusters as determined by the the Dartmouth models of \citet{dotter07}). Table \ref{table:gcparam} lists the age and $t_{rh}$ of each cluster as well as the values of M and $r_h$ used for the calculation of $t_{rh}$.

\begin{table}
  \caption{Globular Cluster Parameters}
  \label{table:gcparam}
  \begin{center}
    \begin{tabular}{lcccc}
      \hline\hline
      {$Name$} &{$M (M_\odot)$} & {$r_h$ (pc)} & {Age (Myr)} &{$t_{rh}$ (Myr)}  \\
      \hline

{M10$^{1,2}$}& {$1.55 \times 10^5$} & {2.3} & {11390} & {743.0} \\
{NGC 5466$^3$}& {$4.47 \times 10^4$} & {13.94} & {13570} & {6224.3}\\
{NGC 6101$^4$}& {$1.45 \times 10^5$} & {9.07} &{12540} & {5311.3} \\
{NGC 6218$^3$}& {$7.24 \times 10^4$} & {4.74} & {12670} & {1504.9}\\
{NGC 6981$^3$}& {$6.46 \times 10^4$} & {7.70} & {10880} & {2971.6}\\

      \hline\hline
      \multicolumn{5}{l}{$^1$ \citet{mclaughlin05}} \\
      \multicolumn{5}{l}{$^2$ \citet{harris96}} \\
      \multicolumn{5}{l}{$^3$ \citet{sollima17}} \\
      \multicolumn{5}{l}{$^4$ \citet{dalessandro15}} \\
    \end{tabular}
  \end{center}
\end{table}

M10 appears to be more segregated than NGC 5466 (more negative $\delta_\alpha$), which is consistent with its shorter present day half-mass relaxation time. Our results are also consistent with \citet{goldsbury13}, who found NGC 5466 was less segregated than M10 based on how concentration varies with stellar mass in each cluster. We caution though that a direct comparison between the two clusters can only be treated as an approximation as the fields of view and mass ranges used to measure $\alpha(r)$ are different. However it should be noted that even though the fields of view used to measure $\delta_\alpha$ are different, since $\alpha(r)$ is approximately linear with $\ln\frac{r}{r_m}$ any effects that field of view have on measuring $\delta_\alpha$ will be minimal. Comparisons of either M10 or NGC 5466 to NGC 6101, NGC 6218, and NGC 6918 cannot be made as the behaviour of $\delta_\alpha$ for stars between 0.5 and 0.8 $M_{\odot}$ is different from that obtained using stars over wider mass ranges.  

Comparing NGC 6101, NGC 6218, and NGC 6918 to each other however is acceptable, as $\delta_\alpha$ in NGC 6101 has been measured over only a slightly wider mass range. NGC 6218 appears to be more segregated than NGC 6981, which is also consistent with their present day relaxation times and with how their concentrations vary with mass \citep{goldsbury13}. Additionally, both of these clusters are significantly more segregated than NGC 6101, consistent with \citet{dalessandro15} where it was stated that NGC 6101 has undergone little to no mass segregation. However, the fact that stars in the mass range observed in NGC 6101 show significantly less segregation than NGC 6981 is surprising as their relaxation times are very similar; to explain the low degree of segregation observed in NGC 6101 \citet{peuten16} recently invoked the presence of a significant population of stellar mass black holes.

\subsection{Comparing Observations to Simulations}\label{s_simresults}

In the following subsections, we compare the evolution of $\delta_\alpha$ in our models to the values measured for M10, NGC 5466, NGC 6101, NGC 6218, and NGC 6981. To ensure a proper comparison, we use the projected half-mass radius of each model cluster at a given time step and assume a mean mass of $\frac{1}{3} M_\odot$ when calculating $t_{rh}$. When measuring the stellar mass function, we only consider model stars that are within the same field of view and mass range as the observed clusters. To restrict the field of view, we found the limits of each observed field in terms of $\frac{r}{r_m}$ and applied the same limits to each simulation at every time-step.

The evolution of $\delta_\alpha$ is compared to both the cluster's dynamical age (traced by the ratio of cluster age to current half-mass relaxation time $\frac{t}{t_{rh}(t)}$) and $\alpha_G$. Globular cluster ages and $t_{rh}$ are listed in Table \ref{table:gcparam}. The uncertainty in each cluster's age is taken from \citet{marinfranch09} and is between 250 and 650 Myr for the clusters studied here. We base our uncertainty in $t_{rh}$ on the recent findings of \citet{shanahan15}, who determined that cluster masses calculated using integrated light estimates may differ from their true values by a factor of two for low metallicity clusters. While only the masses of M10 and NGC 6101 were determined using integrated light profiles, we conservatively apply the same uncertainty to NGC 5466, NGC 6218 and NGC 6981 as well even though their true uncertainties may be lower since their masses were found using multimass dynamical models \citep{sollima17}.

The slope of the global MF $\alpha_G$ has been shown to be a good observational tracer for the fraction of mass lost by a cluster \citep{vesperini97, trenti10, webb15}. As previously discussed, the amount of mass segregation experienced by a cluster should scale with its dynamical age while the amount that $\alpha_G$ has evolved from its initial value will depend on the amount of mass segregation experienced by a cluster. Since a true measure of the global MF for each cluster is not possible, given the restricted fields of view, we are forced to approximate $\alpha_G$ on a cluster to cluster basis. However, we stress here that our choice of how to approximate $\alpha_G$ is relatively inconsequential since the models are given the same treatment as each of the observational datasets. 

For example, for M10, NGC 6218, and NGC 6981 the slope of the mass function of stars with radii within $15\%$ of $r_m$ was used as a tracer of $\alpha_G$, as the MF at $r_m$ is minimally affected by mass segregation and has been shown to be a strong indicator of a clusters global MF \citep[see][]{vesperini97,demarchi00, hurley08}. However, this was not possible for NGC 5466 and NGC 6101 as the observational fields of view do not fully encompass this radial range. For these two clusters we instead used the mass function of all stars within the ACS datasets. While the ACS MF may be slightly top-heavy compared to the actual global MF, as it covers only stars within $r_m$, since the same field of view is used when comparing observations to models then our choice of how to approximate $\alpha_G$ is not biased by projection effects or mass segregation as the effects will be equally present in both measurements.

\subsubsection{M10}

\begin{figure}
\centering
\includegraphics[width=\columnwidth]{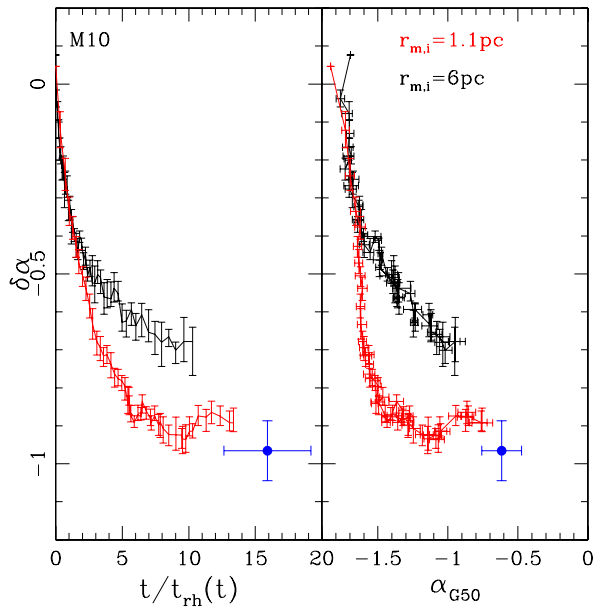}
\caption{Slope of the radial variation in the stellar mass function for stars between 0.3 - 0.8 $M_{\odot}$ as a function of time normalized by current relaxation time (left panel) and the global $\alpha_G$ (right panel) for model clusters with initial masses of $6.3 \times 10^4 M_{\odot}$ and initial half mass radii of 1.1 pc (red) and 6 pc (black). Only model stars within the same field of view as our M10 dataset were used to measure $\delta_\alpha$ and $\alpha_G$. The blue data point represents the observed values for M10.}
  \label{fig:m10}
\end{figure}

Figure \ref{fig:m10} illustrates the evolution of $\delta_\alpha$ with respect to $\frac{t}{t_{rh}(t)}$ and $\alpha_G$ for the two $N$-body model clusters using only stars in the same field of view and mass range as the M10 dataset. The corresponding points for the actual M10 dataset are also plotted. From the left panel of Figure \ref{fig:m10}, it appears that the degree of mass segregation in M10 is consistent with its dynamical age assuming it formed with an initial relaxation time comparable to the $r_{m,i}=1.1$ pc cluster (if not shorter). Since the cluster has undergone a high degree of mass segregation, mass loss has expectedly caused the global MF to evolve from its initial value. We find that $\alpha_G$ for M10 is just a little flatter (less negative) than the two model clusters presented here, implying it has lost a slightly larger fraction of its initial mass. This minor difference can easily be explained either by M10 having a lower initial relaxation time (such that it segregates a bit faster) or M10 experiencing a higher mass loss rate (which allows the global MF to flatten at a faster rate) than the models. While both model clusters have circular orbits at 6 kpc, M10 has an eccentric orbit between 3.4 kpc and 4.9 kpc \citep{dinescu99}. The effective circular orbit \citep{baumgardt03} (the circular orbit distance at which an identical cluster will have the same total lifetime as a cluster with the corresponding eccentric orbit) of M10 is approximately 4 kpc. Hence M10 loses mass at a slightly faster rate than either of the models, but not so much faster that the structural evolution of the cluster is affected which would in turn alter the evolution of $\frac{t}{t_{rh}(t)}$ and $\delta_\alpha$ \citep{webb16}. 

\subsubsection{NGC 5466}

For NGC 5466, we find that the cluster is less segregated given its dynamical age than either of our models would suggest. However, a more significant discrepancy exists between the observed degree of mass segregation in NGC 5466 and its $\alpha_G$. Given a $\delta_\alpha$ of -0.19, the cluster has undergone very little mass segregation such that the mean mass of escaping stars is only slightly less than the mean mass of stars in the entire cluster. With such a low degree of mass segregation, the MF should flatten very slowly as stars escape the cluster. Therefore NGC 5466 should have lost an extremely large amount of mass in order for the MF to reach an $\alpha_G$ of -1.0. If NGC 5466 experienced a significantly higher mass loss rate than our models, such that is structural evolution was strongly affected, then $\delta_\alpha$ would also stop decreasing at an earlier dynamical age once the cluster becomes tidally filling \citep{webb16}. 

\begin{figure}
\centering
\includegraphics[width=\columnwidth]{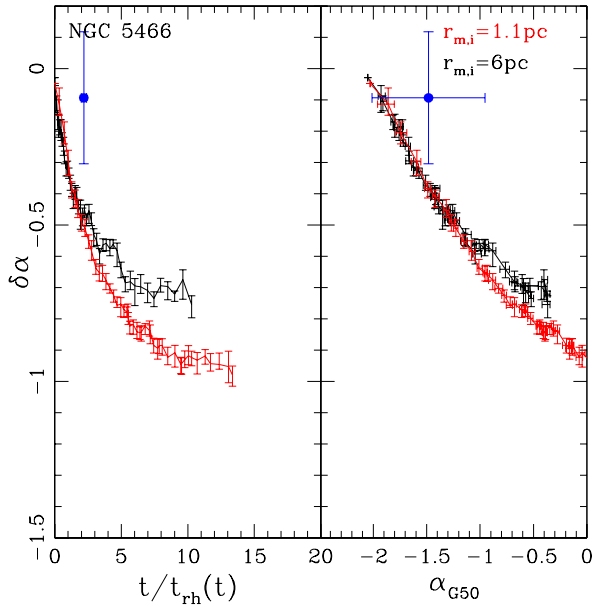}
\caption{Slope of the radial variation in the stellar mass function for stars between 0.4 - 0.8 $M_{\odot}$ as a function of time normalized by current relaxation time (left panel) and the global $\alpha_G$ (right panel) for model clusters with initial masses of $6.3 \times 10^4 M_{\odot}$ and initial half mass radii of 1.1 pc (red) and 6 pc (black). Only model stars within the same field of view as our NGC 5466 dataset were used to measure $\delta_\alpha$ and $\alpha_G$. The blue data point represents the observed values for NGC 5466. Note that the horizontal error bars in the left panel are within the size of the datapoint.}
  \label{fig:ngc5466}
\end{figure}

The fact that NGC 5466 has prominent tidal tails \citep{grillmair06, belokurov06} suggests that it is actively being stripped of stars. However, given that the cluster has an effective circular orbit of 11.8 kpc \citep{dinescu99} it is actually losing mass at a \textit{lower} rate than our model clusters. The tidal tails of NGC 5466 are believed to have formed due to tidal shocks at perigalacticon and disk passages \citep{fellhauer07}. Therefore we instead expect the $\alpha_G$ of NGC 5466 to be closer to its primordial value than the models at a given dynamical age, in disagreement with Figure \ref{fig:ngc5466}.

A second explanation stems from the fact that NGC 5466 may represent a cluster that has been recently accreted by the Milky Way. In fact, a study by \citet{bellazzini03} found that the orbital properties of NGC 5466 suggest the cluster is associated with the Sagittarius dwarf spheroidal galaxy. Hence the tidal field NGC 5466 currently experiences may not be an accurate representation of the mean tidal field the cluster experienced over the majority of its lifetime. The cluster's properties may instead be a result of it forming in a different environment and experiencing a more complex dynamical history. For example, as a member of a dwarf galaxy NGC 5466 could have experienced a much higher mass loss rate than it currently does and may even have experienced a major episode of mass loss when it was accreted by the Milky Way.

\subsubsection{NGC 6101}

\begin{figure}
\centering
\includegraphics[width=\columnwidth]{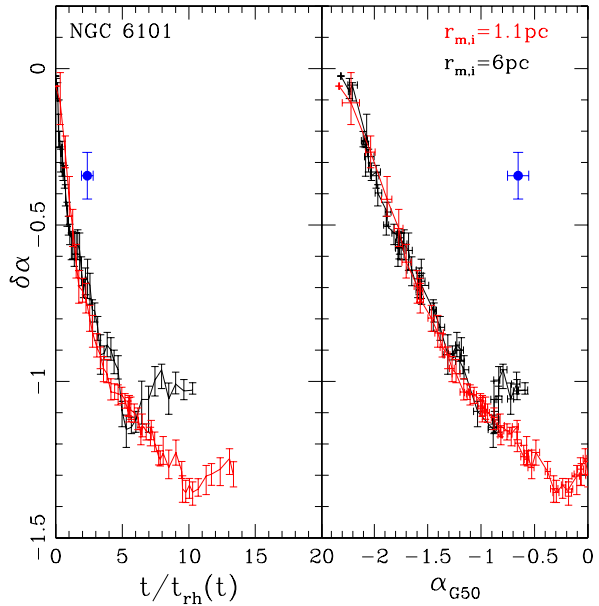}
\caption{Slope of the radial variation in the stellar mass function for stars between 0.5 - 0.8 $M_{\odot}$ as a function of time normalized by current relaxation time (left panel) and the global $\alpha_G$ (right panel) for model clusters with initial masses of $6.3 \times 10^4 M_{\odot}$ and initial half mass radii of 1.1 pc (red) and 6 pc (black). Only model stars within the same field of view as our NGC 6101 dataset were used to measure $\delta_\alpha$ and $\alpha_G$. The blue data point represents the observed values for NGC 6101.}
  \label{fig:ngc6101}
\end{figure}

Similar to NGC 5466, stars in the mass range observed for NGC 6101 are not as segregated as expected for its dynamical age and its mass function is flatter (less negative) than the degree of mass segregation in the cluster should allow (see Figure \ref{fig:ngc6101}). With an effective circular orbit of about 6 kpc \citep{dambis06, balbinot17}, NGC 6101 should lose mass at a similar rate as the model clusters, making the discrepancy between the cluster's dynamical age, $\delta_\alpha$, and $\alpha_G$ even more puzzling. 

As far as the low degree of mass segregation is concerned, a possible solution has been proposed by \citet{peuten16} who recently suggested that a large black hole retention fraction can account for the lack of mass segregation in NGC 6101, building on earlier work by \citet{trenti10} and \citet{lutzgendorf13} (see also recent work by \citet{alessandrini16}). A similar argument could then be made that NGC 5466 also retained a large fraction of stellar mass black holes. However, it is important to emphasize here that although a lack of significant segregation for stars with masses in the range observed can be explained by the presence of a significant population of stellar mass black holes it does not reconcile the discrepancy between $\delta_\alpha$ and $\alpha_G$. Given such a small level of mass segregation, a near-flat global mass function resulting from a preferential loss of low-mass stars can only be reached if the cluster has suffered extreme mass loss (which its current orbit does not support). We will further discuss this point and address the effects that black hole retention has on the coevolution of $\frac{t}{t_{h}(t)}$, $\delta_\alpha$, and $\alpha_G$ in Section \ref{sec:bh}. 

Similar to NGC 5466, NGC 6101 may also represent a cluster that has been recently accreted by the Milky Way. The accretion scenario is supported by the fact that NGC 6101 has a retrograde orbit \citep{geisler95} and that its orbital properties are consistent with the cluster originating in the Canis Major dwarf galaxy \citep{martin04}. Hence NGC 6101 could have had a very complex dynamical history compared to a cluster that has an effective circular orbit in the Milky Way of 6 kpc.

\subsubsection{NGC 6218}

\begin{figure}
\centering
\includegraphics[width=\columnwidth]{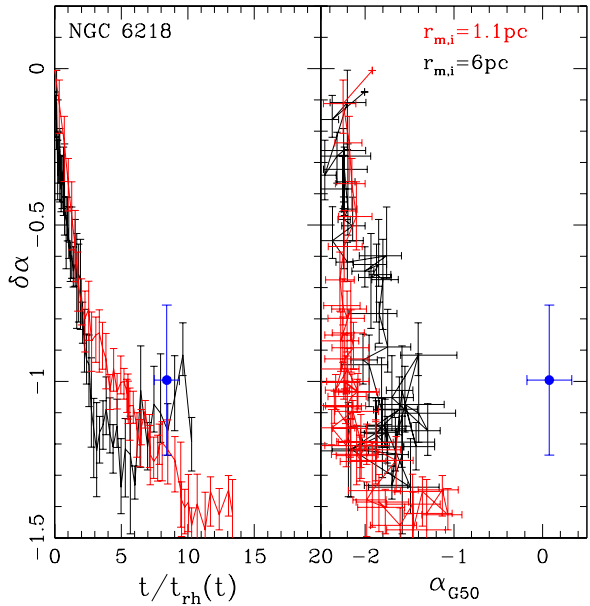}
\caption{Slope of the radial variation in the stellar mass function for stars between 0.55 - 0.8 $M_{\odot}$ as a function of time normalized by current relaxation time (left panel) and the global $\alpha_G$ (right panel) for model clusters with initial masses of $6.3 \times 10^4 M_{\odot}$ and initial half mass radii of 1.1 pc (red) and 6 pc (black). Only model stars within the same field of view as our NGC 6218 dataset were used to measure $\delta_\alpha$ and $\alpha_G$. The blue data point represents the observed values for NGC 6218.}
  \label{fig:ngc6218}
\end{figure}

NGC 6218 represents an interesting case as the degree of mass segregation in the cluster is consistent with its dynamical age when compared to the $r_{m,i}= 6$ pc model (see the left panel of Figure \ref{fig:ngc6218}), but its MF is significantly flatter ($\alpha_G > 0$) than either of the model clusters. Such a flat MF suggests that, if the IMF of this cluster was a Kroupa-like mass function, NGC 6218 must have lost a higher fraction of its initial mass than either of the models. Given the orbital parameters of NGC 6218 from \citet{dinescu99}, the cluster has an effective circular orbit of about 3.4 kpc which would support the idea that the cluster experiences a higher mass loss rate. However, \citet{demarchi06} has found that despite the cluster reaching orbital distances of approximately 3 kpc its orbit would still not yield enough mass loss to produce a near-flat MF. \citet{demarchi06} instead argue that the orbit of NGC 6218 as calculated by \citet{odenkirchen97}, which brings NGC 6218 to a perigalactic distance of 0.6 kpc, is more likely as it would result in a high enough mass loss rate such that the MF of the cluster will be much flatter than the $N$-body models in Figure \ref{fig:ngc6218} within 12 Gyr. In fact, in the \citet{odenkirchen97} scenario NGC 6218 is only 4.5 Gyr away from disruption \citep{demarchi06}.

Unfortunately, since the mass lass rate necessary for resolving the difference between the global MFs of the models and NGC 6218 is so high, the evolution of $\delta_\alpha$ and $\frac{t}{t_{rh}(t)}$ will also be affected. \citet{webb16} found that a higher mass loss rate causes $\delta_\alpha$ to stop decreasing at an earlier dynamical age as tidal stripping removes low mass stars from the cluster faster than they can segregate outwards. Hence a higher mass loss rate cannot simultaneously explain both the cluster's degree of mass segregation and global MF. 

\subsubsection{NGC 6981}

\begin{figure}
\centering
\includegraphics[width=\columnwidth]{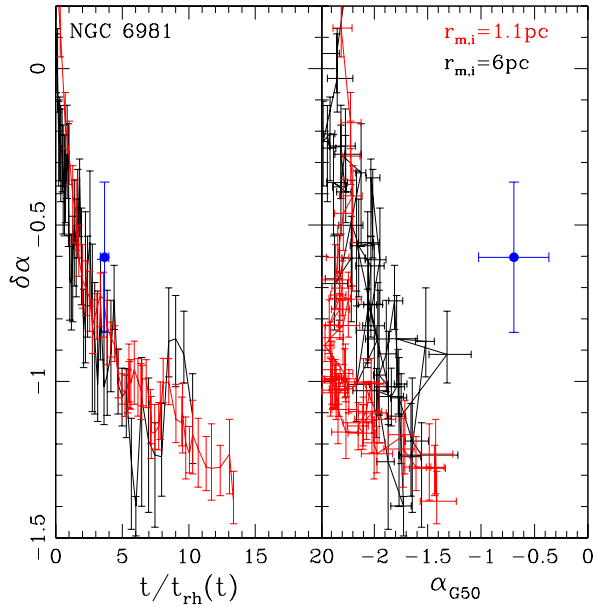}
\caption{Slope of the radial variation in the stellar mass function for stars between 0.55 - 0.8 $M_{\odot}$ as a function of time normalized by current relaxation time (left panel) and the global $\alpha_G$ (right panel) for model clusters with initial masses of $6.3 \times 10^4 M_{\odot}$ and initial half mass radii of 1.1 pc (red) and 6 pc (black). Only model stars within the same field of view as our NGC 6981 dataset were used to measure $\delta_\alpha$ and $\alpha_G$. The blue data point represents the observed values for NGC 6981.}
  \label{fig:ngc6981}
\end{figure}

Finally, the comparison between NGC 6981 and the model clusters is similar to the case of NGC 6218 as the amount of mass segregation in NGC 6981 is in agreement with its dynamical age while the clusters MF is significantly flatter (less negative) than either of the model clusters. However, unlike NGC 6218 the high mass loss rate scenario is not even applicable. NGC 6981 is currently located at a galactocentric distance of 12.9 kpc and estimates of its proper motions indicate it likely does not come within 10 kpc of the Galactic center \citep{dambis06, balbinot17}. Therefore given its dynamical age and the fact that NGC 6981 is only partially mass segregated, its MF should be close to its primordial value. Instead, as shown in Figure \ref{fig:ngc6981}, the MF of NGC 6218 is significantly flatter than what is expected from models assuming a \citet{kroupa93} IMF.

\section{Discussion}\label{s_discussion}

We have compared the evolution of the slope of the global mass function and its variation with clustercentric distance in $N$-body star clusters to the Galactic GCs M10, NGC 5466, NGC 6101, NGC 6218, and NGC 6981. For M10, $\delta_\alpha$ and $\alpha_G$ are consistent with the cluster's dynamical age. NGC 6218 and NGC 6981 on the other hand have $\delta_\alpha$ values that are consistent with their dynamical ages, but have significantly flatter global MFs compared to the $N$-body models. Such a flat MF, if resulting from the preferential loss of low-mass stars, would require the clusters to have suffered a much higher star escape rate. For NGC 6218, its orbit is consistent with the cluster experiencing a higher mass loss rate than the $N$-body models. However, in this scenario the cluster would not reach the same $\delta_\alpha$ as mass segregation stops earlier for clusters that experience high mass loss rates \citep{webb16}. In the case of NGC 6981, its estimated orbit actually yields a lower mass loss rate than the $N$-body models such that a higher mass loss rate is not a viable option. NGC 5466 and NGC 6101 represent the two cases which most strongly disagree with model predictions, as both clusters have undergone very little mass segregation relative to their dynamical ages and have significantly flattened MFs. 

To explain clusters like NGC 6218 and NGC 6981, which have significantly flatter MFs than our models, we first need to explore whether experiencing a high mass loss rate allows for $\alpha_G$ to reach its present day value given the amount of mass segregation currently observed in each cluster. Clusters like NGC 5466 and NGC 6101, which are characterized by a degree of mass segregation for stars in the mass range observed that is less than that predicted by our models, are more difficult to explain. It is important to emphasize that the lack of mass segregation in each observed cluster is not the issue we are focused on, as mechanisms like the effects associated with the presence of a population of stellar mass black holes have been shown to slow the mass segregation of stars \citep{trenti10, lutzgendorf13, webb16, peuten16, alessandrini16}. The key issue instead concerns the coupling between the evolution of $\delta_\alpha$ and the dynamical flattening of the global MF, regardless of the mechanism that is slowing down mass segregation over the stellar mass ranges considered. With little segregation the global mass function can not undergo a significant flattening unless the cluster experiences an extremely high mass loss rate or suffers a major mass loss event. In the case of no segregation, the global MF will not flatten at all no matter what how many stars have escaped the cluster. We therefore need to explore whether higher mass loss rates can even reproduce the MFs of the observed clusters and if slowing the evolution of $\delta_\alpha$ (via black hole retention) will produce clusters with minimal mass segregation and flat MFs.

\subsection{Escaping Stars and the Evolution of $\alpha_G$.}

With the exception of M10, the recurring issue in our comparisons is that the MF of the observed clusters is flatter than the model clusters. A simple explanation would be that these clusters have simply lost a significantly higher fraction of their initial mass than the model clusters they are being compared to. However, what first needs to be determined is whether or not it is even possible for each cluster to reach its present day $\alpha_G$ given its current $\delta_\alpha$ as the evolution of the two parameters is coupled.

To explore whether or not a higher mass loss rate can explain the flatter MFs of certain clusters, we make use of the publicly available code McLuster \citep{kupper11} to generate model clusters with \citet{kroupa93} IMFs over a range of $\delta_\alpha$ values. We specifically setup clusters with primordial mass segregation parameters (S) equal to 0, 0.5, 0.6, 0.7 and 0.73 (where S=0 corresponds to no segregation and S=1 corresponds to extreme segregation). These values of S roughly correspond to initial $\delta_\alpha$ values of 0, -0.25, -0.5, -0.75, and -1.0 when measured using stars between 0.5 and 0.8 $M_\odot$. We then mimic the escape of stars from the cluster by randomly removing stars beyond the cluster's $70\%$ Lagrange radius. 

\begin{figure}
\centering
\includegraphics[width=\columnwidth]{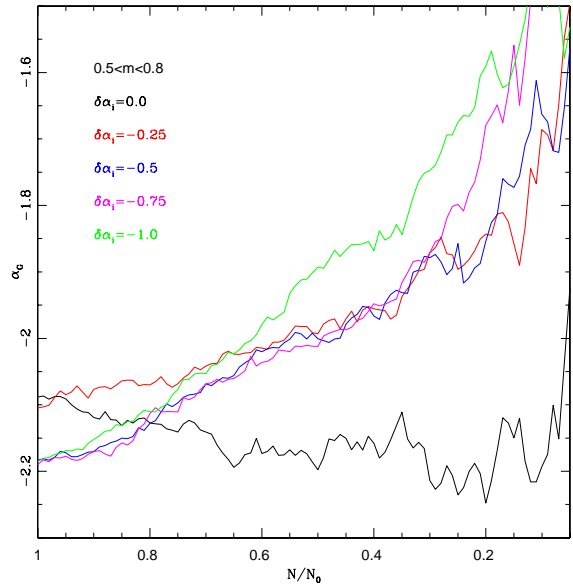}
\caption{Evolution of the global mass function $\alpha$ (measured using stars between 0.5 and 0.8 $M_{\odot}$) as stars beyond the $70\%$ Lagrange radius are randomly removed from model snapshots with a range of $\delta_\alpha$ values.}
  \label{fig:toy}
\end{figure}

We emphasize that this is a simple toy-model, but this experiment essentially illustrates how $\alpha_G$ will evolve as stars escape the cluster under the assumption that the cluster has always had a $\delta_\alpha$ equal to its present day value and that the mass loss rate experienced by the cluster does not affect the minimum $\delta_\alpha$ a cluster can reach. The fraction of stars each cluster must lose to reach a given value of $\alpha_G$ is a lower limit, as the experiment has been optimized to produce the largest possible change in $\alpha_G$ from its initial value. In the more realistic case of $\delta_\alpha$ decreasing from near zero, early mass loss will not have any effect on the global MF while an increased mass loss rate will slow the rate at which $\delta_\alpha$ decreases (which in turn decreases the rate that $\alpha_G$ becomes less negative). The evolution of $\alpha_G$ (measured for stars between 0.5 and 0.8 $M_{\odot}$) with respect to the ratio of the number of stars currently in the cluster to the initial number of stars in the cluster $\frac{N}{N_0}$ is illustrated in Figure \ref{fig:toy}. \par

Figure \ref{fig:toy} reinforces our previous statements that if $\delta_\alpha = 0$, then $\alpha_G$ will not evolve despite stars being able to escape the cluster. Furthermore, for more negative values of $\delta_\alpha$ the escape of stars allows for $\alpha_G$ to increase faster as a function of $\frac{N}{N_0}$. It is interesting to note, however, that none of the cases reach the present day $\alpha_G$ values of NGC 6101, NGC 6218, or  NGC 6981 despite $\alpha_G$ likely evolving at a faster rate in this experiment than in reality. If we consider stars over a wider mass range to compare with NGC 5466, we find that only strongly segregated clusters with initial $\delta_\alpha$ values less than -0.5 reach $\alpha_G = -1$ before dissolution, which is much less than the cluster's present day $\delta_\alpha$ of -0.19. Hence Figure \ref{fig:toy} is revealing that there is a disconnect between the amount of mass segregation in these Galactic clusters and the slope of their present day MFs. These clusters should instead have MFs that are much closer to their primordial values. Therefore, alternative explanations are required to resolve the discrepancy between each cluster's dynamical age, $\delta_\alpha$, and $\alpha_G$.

\subsection{Black Hole Retention}\label{sec:bh}

\begin{figure}
\centering
\includegraphics[width=\columnwidth]{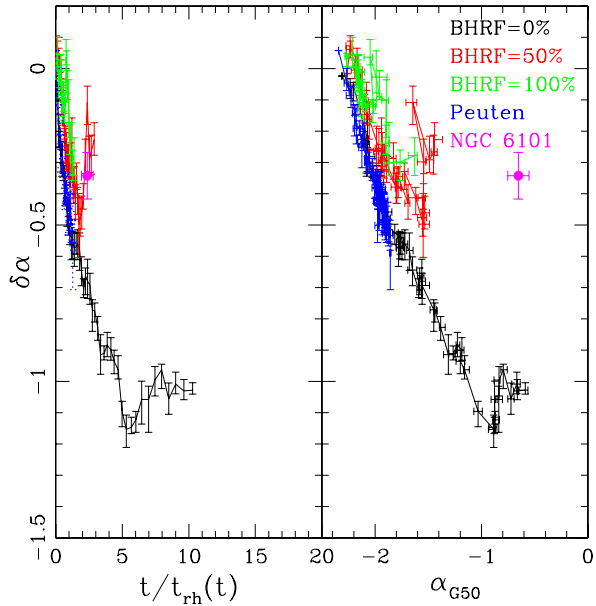}
\caption{Slope of the radial variation in the stellar mass function for stars between 0.5 - 0.8 $M_{\odot}$ as a function of time normalized by current relaxation time (left panel) and the global $\alpha_G$ (right panel) for model clusters with initial masses of $6.3 \times 10^4 M_{\odot}$ and initial half mass radii of 6 pc. Different color lines correspond to model clusters with black hole retention fractions (BHRFs) of $0\%$ (black), $50\%$ (green), and $100\%$ (red). The magenta line corresponds to the model of \citet{peuten16}. Only model stars within the same field of view as our NGC 6101 dataset were used to measure $\delta_\alpha$ and $\alpha_G$. The blue data point represents the observed values for NGC 6101.}
  \label{fig:bh}
\end{figure}

To illustrate how mechanisms which slow the evolution of $\delta_\alpha$ affect the relationship between the degree of mass segregation in a cluster and MF flattening, we have re-simulated the $r_{m,i}=6$ pc clusters but with black hole retention fractions of $50\%$ and $100\%$. The new simulations are compared to the original models in Figure \ref{fig:bh}, given the observational constraints of NGC 6101 as it represents the largest discrepancy with the models. It should be noted that the new simulations have only been evolved until they lose the same amount of mass as the original models which retain no black holes. For comparison purposes, we have also included in Figure \ref{fig:bh} the model from \citet{peuten16} that the authors found to best reproduce the degree of mass segregation in NGC 6101. The \citet{peuten16} model has an initial mass of $6.3 \times 10^4 M_{\odot}$, an initial half-mass radius of 7.6 pc and a $100\%$ black hole retention fraction. While less massive and smaller than the true NGC 6101 progenitor, these initial conditions ensure the $N$-body model has the same initial half-mass relaxation time as the most likely NGC 6101 progenitor the authors find using EMACSS \citep{alexander14}. 

In agreement with previous studies, and as discussed in detail in \citet{webb16}, the retention of black holes expectedly causes $\delta_\alpha$ to stop decreasing at an early $\frac{t}{t_{rh}(t)}$. In fact, each of the model clusters with $100\%$ retention fractions reach similar final values of $\delta_\alpha$ as NGC 6101. However, both models have shorter $\frac{t}{t_{rh}(t)}$ ratios and have significantly steeper MFs than NGC 6101. Retaining black holes results in clusters having larger cores and expanding to larger half-mass radii \citep{webb16}, resulting in the cluster having longer core and half-mass relaxation times. Furthermore, since model clusters that retain black holes are dynamically younger and less mass segregated, the escape of stars causes the global MF to evolve quite slowly (as seen in Figure \ref{fig:toy}). In fact, $\alpha_G$ in models that retain black holes is even steeper than $\alpha_G$ in the modes that do not. Hence, while retaining black holes can result in clusters being less segregated and having less negative values of $\delta_\alpha$, the discrepancy between the observed slope of the MF and the one expected, if a standard IMF is adopted, remains.

\subsection{Alternative Explanations}

With neither higher mass loss rates or mechanisms for slowing mass segregation (e.g. back hole retention) accounting for the discrepancy between the dynamical ages, degree of mass segregation, and MFs of NGC 5466, NGC 6101, NGC 6218, and NGC 6981, we consider in the following sections two additional factors which were found to affect the co-evolution of $\delta_\alpha$, $\frac{t}{t_{rh}}$, and $\alpha_G$ in \citet{webb16}. 

\subsubsection{Primordial Mass Segregation}

Assuming that a cluster can reach its final $\delta_\alpha$ before the global MF has evolved, as we did in the previous section, is similar to assuming the cluster formed primordially mass segregated. A key difference, however, is that if a cluster forms primordially mass segregated then the structural evolution of the cluster will be very different than the non-primordially segregated case. Primordially mass segregated clusters have been shown to undergo a large initial expansion as stars lose mass via stellar evolution such that they will lose additional mass via tidal stripping \citep[e.g.][]{baumgardt08, vesperini09, haghi14, haghi15}. With respect to the co-evolution of $\delta_\alpha$, $\frac{t}{t_{rh}}$, and $\alpha_G$, \citet{webb16} found that primordially mass segregated clusters would reach similar final values of $\delta_\alpha$ and $\frac{t}{t_{rh}}$ as non-primordially mass segregated clusters, but have significantly more evolved MFs. In some cases, the initial $\delta_\alpha$ was less than the final $\delta_\alpha$ such that early mass loss would cause $\alpha_G$ to evolve more than if the cluster always had its final $\delta_\alpha$ (as we explored in Figure \ref{fig:toy}).

In a recent comparison between the observed $\alpha(r)$ in Pal 4 and $N$-body star cluster simulations, \citet{zonoozi17} was in fact able to reproduce the observed radial variation in Pal 4's MF by assuming the cluster formed highly primordially mass segregated and has an eccentric orbit. The authors required a mass segregation parameter S equal to 0.9 to match the models to observations. Both assumptions were necessary to explain how such a dynamically young ($\frac{t}{t_{rh}(t)} \sim 4$) cluster could have significant radial variation in $\alpha(r)$ and lose enough mass to have an evolved MF ($\alpha_G = -1.14 \pm 0.25$ for stars between 0.55 and 0.85 $M_{\odot}$).

\begin{figure}
\centering
\includegraphics[width=\columnwidth]{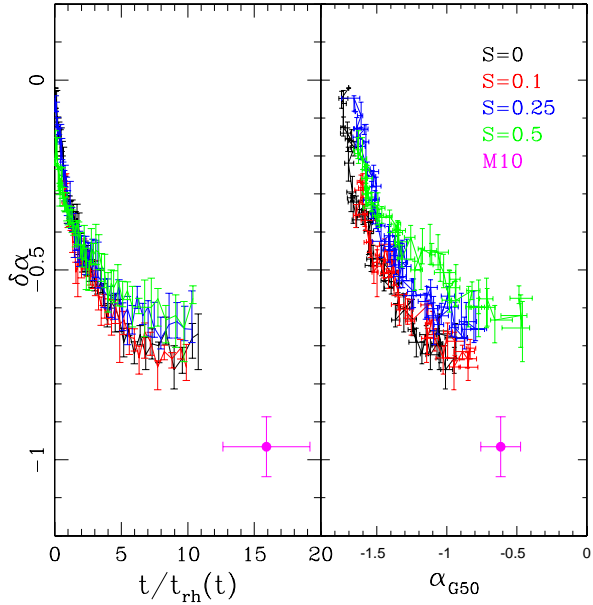}
\caption{Slope of the radial variation in the stellar mass function for stars between 0.3 - 0.8 $M_{\odot}$ as a function of time normalized by current relaxation time (left panel) and the global $\alpha_G$ (right panel) for a non-primordially mass segregated model cluster with an initial mass of $6.3 \times 10^4 M_{\odot}$ and an initial half mass radius of 6 pc (black). Primordially mass segregated versions of the cluster are illustrated in red (S=0.1), blue (S=0.25) and green (S=0.5). Only model stars within the same field of view as our M10 dataset were used to measure $\delta_\alpha$ and $\alpha_G$. The magenta data point represents the observed values for M10.}
  \label{fig:PMS}
\end{figure}

To explore the effects of mass segregation on our results, we make use of simulations of the $r_{m,i} = 6$ pc cluster (which normally have S=0) from \citet{webb16} that have S = 0.1, 0.25, and 0.5 (again set up using the publicly available code McLuster \citep{kupper11}). Since the effects of mass segregation are more visible when the mass function is measured over a wide mass range, we plot the evolution of $\delta_\alpha$ with respect to $\frac{t}{t_{rh}(t)}$ and $\alpha_G$ given the observational constraints of M10 in Figure \ref{fig:PMS}. As expected, early mass loss causes the MF to evolve quicker for the  primordially mass segregated cases, with the S=0.5 cluster surpassing the $\alpha_G$ of M10. Hence primordial mass segregation may offer an explanation for the $\alpha_G$ of clusters like NGC 6218 and NGC 6981, which show the appropriate degree of mass segregation given their $\frac{t}{t_{rh}(t)}$ but not enough to explain their MF. However, the degree of primordial mass segregation would have to be very high in order for stars within such a narrow mass range (0.55-0.8 $M_\odot$) to be affected. For clusters like NGC 5466 and NGC 6101, which have evolved MFs despite undergoing very little mass segregation, the disconnect between their $\delta_\alpha$ and $\alpha_G$ cannot be explained by primordial mass segregation alone.

 \subsubsection{A Non-Universal IMF}

Finally, one remaining explanation for why the dynamical age, $\delta_\alpha$, and $\alpha_G$ of select clusters are in disagreement could be that the stellar IMF is not universal. \citet{webb16} demonstrated that the evolution of $\delta_\alpha$ is unaffected by the functional form of the IMF. Hence if clusters instead form with an initially flatter IMF (compared to a \citet{kroupa93} IMF or \citet{kroupa01} IMF) than the evolution of $\delta_\alpha$ with respect to $\alpha_G$ in the right panels Figures \ref{fig:m10} to \ref{fig:ngc6981} will begin at a flatter (less negative) $\alpha_G$ and can more easily reach with the present day $\alpha_G$ values of each cluster. In fact a non-universal IMF may offer the only explanation for the MFs of NGC 5466 and NGC 6101, considering how little mass segregation they have undergone, such that their present day MFs are close to their primordial values. With the orbits of NGC 5466 and NGC 6101 suggesting they were accreted from a dwarf galaxy \citep{geisler95, bellazzini03, martin04}, it is possible that their individual formation environments resulted in clusters forming with unique IMFs. However, there is currently no consensus in the literature regarding whether or not the stellar IMF is universal. Various studies have used the stellar populations of early type galaxies \citep[e.g.][]{cappellari12,conroy13,vandokkum16,coulter17,peacock17} and globular clusters  \citep[e.g.][]{mcclure86,bastian10,strader09,marks14,zaritsky14,shanahan15} to argue for and against universality. Additional wide-field studies of Galactic GCs, which allow for both $\delta_\alpha$ and $\alpha_G$ to be accurately measured, will be able to shed further light on this fundamental issue.

\section{Conclusion}\label{s_conclusion}

We have measured and compared the degrees of mass segregation (traced by $\delta_\alpha$) and the global MFs of five Galactic globular clusters to direct $N$-body star cluster simulations. Three of the clusters in this study (M10, NGC 6218, and NGC 6981) all demonstrate the appropriate amount of mass segregation given their dynamical ages. The global MF of M10 is also in agreement with the degree of mass segregation in the cluster and the mass loss rate it experiences on its current orbit. However the MFs of NGC 6218 and NGC 6981 are flatter than our models, which would indicate they have lost a higher fraction of their initial mass. The other two clusters that we consider (NGC 5466 and NGC 6101) are in strong disagreement with our models, as they show little segregation despite their dynamical ages and have relatively flat MFs. Clusters that have undergone very little mass segregation should instead have MFs that are near-primordial as the escape of stars from the cluster is independent of stellar mass.

We explore whether higher mass loss rates, slowing the evolution of $\delta_\alpha$ via black hole retention, primordial mass segregation, or a non-Universal IMF could resolve any of the discrepancies between the observed clusters and the $N$-body models. While a higher mass-loss rate does have the effect of increasing the rate at which the $\alpha_G$ flattens, we find that higher mass loss rates do not result in $\alpha_G$ evolving enough from its initial value to match each cluster's present day $\alpha_G$. For NGC 6218 and NGC 6981 in particular, taking into consideration that a higher mass loss rate will also slow the mass segregation rate in a cluster, a higher mass loss rate would also yield a discrepancy between the dynamical ages and the amount of mass segregation in each cluster.

Slowing the evolution of $\delta_\alpha$, specifically via the retention of black holes, has been used to explain the lack of mass segregation in NGC 6101 \citep{peuten16}. However, our models show that being able to explain the lack of mass segregation fails to resolve the fact that clusters like NGC 5466 and NGC 6101 have near-flat global MFs. In fact, slowing the evolution of $\delta_\alpha$ using black hole retention causes the MF to evolve even more slowly as stars escape the cluster compared to models which retain no black holes.

In two cases, mainly NGC 6218 and NGC 6981, primordial mass segregation offers a potential explanation for their flat MFs. When $\delta_\alpha$ is already in agreement with the cluster's dynamical age, having the cluster form primordially mass segregated allows for stars which escape the cluster at early times to be preferentially lower in mass, resulting in clusters having flatter MFs after 12 Gyr. An extremely high degree of primordial mass segregation was recently suggested by \citet{zonoozi17} to explain radial variation in the MF of Pal 4.

Our comparison between $N$-body simulations and observed globular clusters has, however, revealed that not all clusters follow the predicted co-evolution of $\frac{t}{t_{rh}(t)}$, $\delta_\alpha$, and $\alpha_G$. Certain clusters, specifically NGC 5466 and NGC 6101, have proven difficult to model as their MFs are flatter than their current degree of mass segregation should allow. A unique formation scenario (e.g. a non-universal IMF) and/or a complex dynamical history (e.g. accretion) is likely required to explain these two clusters. Additional wide-field studies of Galactic globular clusters will help determine whether $\frac{t}{t_{rh}(t)}$, $\delta_\alpha$, and $\alpha_G$ are more commonly in agreement with our theoretical predictions or whether a more complex dynamical history and, possibly, variations in the IMF need to be invoked to explain the observed properties of the cluster mass function.

\section*{Acknowledgements}

This work was made possible in part by the facilities of the Shared Hierarchical Academic Research Computing Network (SHARCNET:www.sharcnet.ca) and Compute/Calcul Canada, in part by Lilly Endowment, Inc., through its support for the Indiana University Pervasive Technology Institute, and in part by the Indiana METACyt Initiative. The Indiana METACyt Initiative at IU is also supported in part by Lilly Endowment, Inc. The authors would also like to thank Mark Gieles, Miklos Peuten, and Eduardo Balbinot for helpful discussions regarding their recent works.

\bsp

\label{lastpage}

\end{document}